\documentclass[12pt]{article}
\usepackage{amsfonts}
\usepackage{amssymb}
\usepackage{cite}
\def\hybrid{\topmargin -20pt    \oddsidemargin 0pt
        \headheight 0pt \headsep 0pt
        \textwidth 6.25in       
        \textheight 9.5in       
        \marginparwidth .875in
        \parskip 5pt plus 1pt   \jot = 1.5ex}

\hybrid
\newcommand{\beqn}{\begin{eqnarray}}
\newcommand{\eeqn}{\end{eqnarray}}
\newcommand{\be}{\begin{equation}}
\newcommand{\ee}{\end{equation}}
\newcommand{\non}{\nonumber \\}
\newcommand{\vol}{{\cal V}}

\newcommand{\pu}{\partial_{\mu}}
\newcommand{\po}{\partial^{\mu}}

\newcommand{\tw}{\tilde W}

\def\wf{\Delta}
\begin {document}
\begin{titlepage}
\begin{center}

\hfill hep-th/0103068\\
\vskip 2cm
{\large \bf M-theory compactified on Calabi-Yau fourfolds with 
background flux}\footnote{Work 
supported by:
DFG -- The German Science Foundation,
GIF -- the German--Israeli
Foundation for Scientific Research and the
DAAD -- the German Academic Exchange Service.}

\vskip .5in

{\bf Michael Haack\footnote{email: {\tt
haack@physik.uni-halle.de}}
and Jan Louis\footnote{email: {\tt j.louis@physik.uni-halle.de}}}  \\

\vskip 0.8cm
{\em Fachbereich Physik, Martin-Luther-Universit\"at Halle-Wittenberg,\\
Friedemann-Bach-Platz 6, D-06099 Halle, Germany}

\end{center}

\vskip 2.5cm

\begin{center} {\bf ABSTRACT } \end{center}

We perform the Kaluza-Klein reduction of M-theory 
on warped Calabi-Yau fourfolds with non-trivial
four-form flux turned on. The resulting scalar- 
and superpotential
is computed and compared with the superpotential obtained
by Gukov, Vafa and Witten using different methods.

\vskip 1cm

\vfill

March 2001
\vskip 1cm

\end{titlepage}

String vacua with 4 unbroken supercharges are of particular interest
due to their possible phenomenological properties.
In four space-time dimensions such vacua  have 
traditionally been constructed 
as compactifications of the heterotic string on Calabi-Yau
threefolds $Y_3$ or conformal field theoretic generalizations thereof.
More recently other classes of vacua with four supercharges
have been considered
such as F-theory compactified on elliptically
fibered Calabi-Yau fourfolds  $Y_4$ \cite{V}.

In this paper we study 
M-theory compactified on $Y_4$ whose low energy
effective theory has three space-time dimensions 
and also four unbroken supercharges.
Aspects of such compactifications have 
been considered in refs.\ \cite{BB,W,PM,KLRY,DRS,HL,GSS,GVW} 
and in an appropriate limit they are related
to F-theory on $Y_4$ \cite{V}.
In a previous paper \cite{HL} we derived the low energy
effective Lagrangian in the large volume limit 
for the class of fourfolds with vanishing Euler number $\chi=0$.
In particular this constraint implied that the metric is a direct
product $M_3\times Y_4$ and no background flux had to be 
turned on. In this paper we generalize the previous analysis
to the case $\chi\neq 0$ which requires either space-time filling 
membranes or a non-trivial
four-form flux $F_4$ on $Y_4$ and a warped 
space-time metric \cite{BB}.
We show that a non-trivial $F_4$ introduces 
Chern-Simons terms and a potential which in turn
can be derived from two superpotentials. 
The presence of the Chern-Simons terms  has 
first been noticed in ref.\ \cite{PM}
while the superpotentials have been 
proposed in \cite{GVW}. 
We find basic agreement with ref.\ \cite{GVW}
except that one of our superpotentials is the 
real version of the corresponding superpotential 
given in \cite{GVW}
.\footnote{
There is another way to generate a
superpotential in three dimensions 
by wrapping $5$-branes over certain $6$-cycles of
$Y_4$ \cite{W}. This can however not occur if 
there is a non-vanishing
four-form flux localized on a four-dimensional submanifold of the $6$-cycle
\cite{DMW}. Such contributions are not consider here.} 
Similar superpotentials arise in the compactification of type IIA
on $Y_4$ \cite{G} and their properties have been 
studied in refs.\
\cite{GVW,G,GGW,HLM}.\footnote{%
An expanded version of our paper
can be found in \cite{MH}.}


Let us first briefly recall the compactification of
M-theory on a fourfold $Y_4$ with $\chi(Y_4)=0$.
For this situation the low energy effective Lagrangian
in the large volume limit has been derived in ref.\ \cite{HL}.
The starting point is 
the eleven-dimensional supergravity which only contains
a metric and a three-form $A_3$ (with field strength
$F_4 = dA_3$)
as massless bosonic components.
The 11-dimensional supergravity action is given by \cite{CJS}
\be\label{Sstart}
{\cal S} =  \frac{1}{2 \kappa_{11}^2} \int d^{11} x \sqrt{-\hat g} 
\left[ \hat R - \frac{1}{2} |F_{4}|^{2} \right] - \frac{1}{12 \kappa_{11}^2}
\int A_{3} \wedge F_{4} \wedge F_{4}\ .
\ee
When compactified on a Calabi-Yau fourfold
additional scalar and vector fields appear in the
resulting three-dimensional effective action.
{}From the metric $h^{1,1}$ real scalar fields 
$M^A, A=1,\ldots,h^{1,1},$ 
which parameterize the deformation of the K\"ahler form
and $h^{3,1}$ complex scalar fields 
$Z^\alpha, \alpha=1,\ldots,h^{3,1},$
which parameterize the deformation of the complex structure
arise.
In addition the three-form $A_3$ leads to $h^{1,1}$ 
vector fields $A_\mu^A$ and $h^{2,1}$ complex scalar
fields $N^I, I=1,\ldots,h^{2,1}$.
Together these fields form $h^{2,1}+h^{3,1}$ chiral multiplets
and $h^{1,1}$ vector multiplets.
For simplicity we consider only the case
where the $(2,1)$-scalars $N^I$ are frozen and 
perform a Kaluza-Klein reduction keeping only the
massless $(1,1)$ and $(3,1)$ modes.  In the large volume
limit this results in \cite{HL}
\beqn\label{M3d}
{\cal L}^{(3)}_0 & = & \frac{1}{2} R^{(3)} 
- G_{\alpha \bar{\beta}} \pu Z^\alpha 
\po \bar{Z}^{\bar{\beta}} 
- \frac{1}{4}\, \vol^2 G_{AB} 
F^A_{\mu \nu} F^{B \mu \nu}\\
&&-\, \frac{1}{2}\, G_{AB} \pu M^A \po M^B  
-\ \frac{1}{2}\, \pu \ln \vol\, \po \ln \vol  \ .\nonumber
\eeqn
$G_{\alpha \bar{\beta}}$ is the K\"ahler metric
on the space of $(3,1)$-forms given by \cite{AS}
\be \label{Galphabeta}
G_{\alpha \bar{\beta}}\  =\
- e^{K_{3,1}}
\int_{Y_4} \Phi_{\alpha} \wedge \bar{\Phi}_{\bar{\beta}}
\ =\ \partial_{\alpha} 
\bar\partial_{{\bar \beta}} 
K_{3,1} \ , \qquad
K_{3,1} =
-\ln \Big[ \int_{Y_4} \Omega \wedge 
\bar{\Omega} \Big]\ ,
\ee
where
$\Omega $ is the unique holomorphic
$(4,0)$-form on $Y_4$, $\bar\Omega$ its complex
conjugate and  $\Phi_{\alpha}$ a basis of $H^{3,1}(Y_4)$.
$G_{AB}$ is the metric 
on the space of $(1,1)$-forms defined as \cite{S}
\be \label{GAB}
G_{AB} = \frac{1}{2 {\vol}} 
\int_{Y_4} e_{A} \wedge 
\star e_{B} = -\frac12 \partial_A\partial_B \ln\vol\ ,
\ee
where
$\vol$ is the volume of $Y_4$
\be
{\vol} = 
\int_{Y_4} d^{8} \xi \sqrt{{g}} =
 \frac{1}{4!} 
\int_{Y_4} {J} \wedge {J} 
\wedge {J} \wedge {J} = \frac{1}{4!} 
d_{ABCD} {M}^A 
{M}^B {M}^C {M}^D\ .
\label{eqvolumekf}
\ee
$J = M^A  e_A$ is the K\"ahler form 
and 
$d_{ABCD} = \int_{Y_4} e_A 
\wedge e_B \wedge e_C \wedge e_D$ are the classical 
intersection numbers of $Y_4$. 

The Lagrangian (\ref{M3d}) can also be displayed in dual
variables where one dualizes the vectors $A_\mu^A$ 
to scalars $P^{A}$ and defines the complex K\"ahler coordinates
\be
T^{A} = \frac{1}{\sqrt{8}} 
( i P^{A} + \vol G_{AB} M^{B} )\ .
\ee
In this basis the Lagrangian (\ref{M3d}) reads \cite{HL}
\be\label{expect}
{\cal L}^{(3)}_0 =  \frac{1}{2} R^{(3)} 
- G_{\alpha \bar{\beta}}\, \pu Z^\alpha \po \bar{Z}^{\bar{\beta}} 
-\,  G_{A\bar B}^K\, \pu T^A \po \bar T^{\bar B}\ , 
\ee
where the metric $ G_{A\bar B}^K$ is K\"ahler and given by
\be\label{K3def}
G_{A\bar B}^K  =\partial_{A}\bar\partial_{\bar B}
 K_{1,1} \ , \qquad
K_{1,1} =  - 3 \ln \vol = 
-\ln\Big[(T^A+\bar T^A) \vol G^{-1}_{AB} (T^B+\bar T^B)\Big]\ .
\ee

Now we come to the main objective of the paper,
that is generalize the previous analysis to the case where
$\chi\neq 0$. 
In the absence of space-time 
filling membranes an $A_3$-tadpole can only be avoided
in the  $D=3$ theory if a 
non-trivial four-form flux $F_4$ along
the internal Calabi-Yau fourfold is turned on \cite{BB,SVW,DM}.
This arises due to the presence of the  higher order term
(in $\kappa_{11}$) \cite{DLM}
\be \label{r4action1}
\delta {\cal S}_1 =
- T_2 \int A_3 \wedge X_8\ , 
\ee
where $T_2 \equiv (2 \pi)^{2/3} 
(2 \kappa_{11}^2)^{-1/3}$ and
\be\label{X8}
X_8 = \frac{1}{192\,(2 \pi)^4} 
\left[ \mbox{tr} \hat R^4  -\frac{1}{4} 
(\mbox{tr} \hat R^2)^2 \right]\ , \qquad 
\int_{Y_4} X_8\ =\ -{\chi\over 24}\ .
\ee
Combining eqs.\ (\ref{Sstart}), (\ref{r4action1}), (\ref{X8})
results in absence of space-time filling membranes 
in the consistency condition  \cite{BB,SVW,DM}
\begin{equation} \label{tadpol}
\frac{1}{4 \kappa^2_{11}} \int_{Y_4} F_4 \wedge F_4 \
=\ \frac{T_2}{24}\ \chi\ ,
\end{equation}
which requires a non-vanishing (internal) $F_4$ if $\chi\neq0$.
Furthermore,
supersymmetry can only be maintained if the metric is
not a direct product $M_3\times Y_4$ but instead 
includes a warp factor $\wf$ \cite{BB,S2,dWS}
\begin{equation} \label{warpmetric}
\hat g_{MN} = \left( \begin{array}{cc}
                     e^{-\wf} g_{\mu \nu}& 0 \\
                     0 & e^{\frac{1}{2} \wf} g_{mn}
                     \end{array} \right)\ ,
\end{equation}
where  to leading order in $\kappa_{11}$ $g_{mn}$ is
a Ricci-flat Calabi-Yau metric. The warp factor $\wf$
obeys \cite{BB}
\begin{equation} \label{laplacephi}
\nabla_m \partial^m 
 \wf = -\frac{1}{3} \star (F_4 \wedge F_4) 
- \frac{4}{3} T_2 \kappa_{11}^2
\star X_8\ , 
\end{equation}
where the 
Laplace operator and the Hodge 
$\star$-operator are 
defined with respect to the metric $g_{mn}$.\footnote{We have adapted 
the formula given in \cite{BB} to our conventions.}
Thus for compactifications
with $\chi\neq0$ higher order terms have to be taken into
account and they in turn warp the three-dimensional
space-time metric.

The Kaluza-Klein reduction is a good approximation
if the size of the internal $Y_4$ manifold is large
compared to the 11-dimensional Planck length 
$ l_{11}^9 = \kappa_{11}^2$
or in other words for $l_Y \gg l_{11}$ where
$l_Y^8$ is the `average' size of $Y_4$.
{}From eq.\ (\ref{tadpol}) we infer that 
$F_4 \sim {\cal O} (l_{11}^3/l_Y^4)$
while eq.\ 
(\ref{laplacephi}) implies
$\wf \sim{\cal O} (l_{11}^6/l_Y^6)$ 
so that in the limit $l_{11}/l_Y \to 0$ 
the metric (\ref{warpmetric}) 
becomes the unwarped product metric 
and $F_4$ vanishes \cite{GSS}.\footnote{Strictly speaking 
$\Delta$ could have a harmonic part. However, this should 
also vanish in the limit $l_{11}/l_Y \to 0$ in order to ensure
that (\ref{warpmetric}) becomes the unwarped metric.}

In this paper we focus on the
three-dimensional effective theory with at most
two derivatives and compute some of the corrections
to the Lagrangian (\ref{M3d}) which result from 
higher order terms. Specifically we compute
the Chern-Simons term and the potential to order 
${\cal O} (\kappa_{11}^{-2/3})$ while 
the corrections to the kinetic terms of (\ref{M3d})
are not calculated. With this restriction
the only other 11-dimensional term
we need to consider is\footnote{All other
bosonic higher derivative terms 
which are related via supersymmetry to the ones given in 
(\ref{r4action1}) and (\ref{r4action})  
are proportional to the Ricci-tensor or contain at least
one $4$-form field strength \cite{T}. Their contribution to the potential is 
therefore subleading.} 
\be \label{r4action}
\delta {\cal S}_2 =
b_1 T_2 \int d^{11} x \sqrt{-g} (J_0 - \frac{1}{2} E_8)
\ , 
\ee
where $b_1^{-1}\equiv (2 \pi)^4 3^2 2^{13}$ and 
\beqn \label{e8j0}
E_8 & = & \frac{1}{3!}
\epsilon^{ABCM_1 N_1 \ldots M_4 N_4}
\epsilon_{ABCM_1' N_1' \ldots M_4' N_4'}
\hat R^{M_1' N_1'}\! _{M_1 N_1} 
\ldots \hat R^{M_4' N_4'}\! _{M_4 N_4}\ , \\
J_0 & = & t^{M_1 N_1 \ldots M_4 N_4} 
t_{M_1' N_1' \ldots M_4' N_4'} \hat R^{M_1' N_1'}\! _{M_1 N_1}
\ldots \hat R^{M_4' N_4'}\! _{M_4 N_4} + \frac{1}{4} E_8\ 
.\nonumber
\eeqn
The tensor $t$ is defined by
$t^{M_1 \ldots M_8} A_{M_1 M_2} \ldots A_{M_7 M_8}  
  = 24 {\rm tr} A^4 - 6 ({\rm tr} A^2)^2$
for antisymmetric tensors $A$.\footnote{We follow here the conventions of \cite{T} which differ from the tensor 
$t_8$ used in \cite{GSW} in that the
$\epsilon$-term is omitted.} 
Note that $E_8$ given in (\ref{e8j0}) 
 is not the eight-dimensional Euler density but an
11-dimensional
generalization of it. 
More generally one can define \cite{T}
\beqn \label{euler3}
E_n (M_D) & = & \frac{1}{(D-n)!}\, \epsilon_{N_1 \ldots N_{D-n} N_{D-n+1} 
\ldots N_D} 
\epsilon^{N_1 \ldots N_{D-n} N'_{D-n+1} \ldots 
N'_D} \non
& & R^{N_{D-n+1} N_{D-n+2}}\, \! _{N'_{D-n+1} N'_{D-n+2}} 
\ldots R^{N_{D-1} N_D}\, \! _{N'_{D-1} N'_D}\ ,
\eeqn 
where $D$ denotes the real dimension of the manifold. 
Then $E_8(Y_4)$ is proportional to the
 eight-dimensional Euler density, i.e.
$ 12b_1 \int_{Y_4} d^8 \xi \sqrt{g}\, E_8(Y_4)  = \chi$
holds.

The next step is to  reduce the action ${\cal S} +
\delta {\cal S}_1 + \delta {\cal S}_2$ consisting of the terms
given in (\ref{Sstart}), (\ref{r4action1}) 
and (\ref{r4action}).
Following ref.\ \cite{BB} we suppose that the only non-vanishing 
components of $F_4$ are $F_{mnpq}$ and 
$F_{\mu \nu \rho m}$ where the latter  are related via
supersymmetry to the warp factor via
$F_{\mu \nu \rho m} = \epsilon_{\mu \nu \rho} 
\partial_m e^{-\frac{3}{2} \wf}$.
In the reduction of the 11-dimensional 
Einstein-Hilbert term one obtains the three-dimensional
Einstein-Hilbert term $R^{(3)}$ and an additional
term proportional to $\int_{Y_4}\wf\nabla_m \partial^m \wf$
which is of higher order.\footnote{The absence
of possible other terms in the presence of a $4$-form flux is 
carefully discussed in ref.\ \cite{MH}.}
Furthermore,
the integral $\int_{Y_4} d^8 \xi \sqrt{g^{(8)}} J_0$ vanishes 
for Ricci-flat K\"ahler manifolds \cite{GW,FPSS}.\footnote{%
$J_0$ is the sum of an internal and an external part. 
Since $J_0$ can be expressed through the Weyl-tensor 
only \cite{BG,GKT} 
the external part vanishes because the Weyl tensor  
vanishes identically in $D=3$.}  
Again, at leading order we can assume the metric 
to be Ricci flat and neglect the effect of the 
warp factor in $J_0$ as a higher order contribution.
To evaluate the leading order contribution
to $E_8$ we use the fact that 
on a product space $M_3 \times Y_4$ one has 
$E_8 (M_3 \times Y_4) = - E_8 (Y_4) + 4 E_2 (M_3) E_6 (Y_4)$
where $E_2 (M_3) = -2R^{(3)}$ holds.

The details of the Kaluza-Klein reduction procedure
can be found in \cite{HL,MH} while here we only give 
a few intermediate steps.
First of all one obtains a non-canonical
Einstein term  in  the three-dimensional
effective action 
\be \label{r4d3actioni}
{\cal S}^{(3)}  =  \frac{1}{2\kappa_{11}^2} 
\int d^3 x \sqrt{-g^{(3)}} \Lambda  R^{(3)}  
+ \ldots \ ,
\ee
where
\be 
\Lambda  = 
 \vol_\wf + 8 \kappa_{11}^2 b_1 T_2 
\int_{Y_4} d^8 \xi \sqrt{g^{(8)}} E_6(Y_4)\ ,
\ee
and $\vol_\wf = \int_{Y_4} d^{8} \xi e^{-\Delta/2} \sqrt{\hat g^{(8)}}$ 
denotes a warped Calabi-Yau volume, where $\hat g^{(8)}$ 
is the internal part of  
the warped metric (\ref{warpmetric}).
With the help of a Weyl rescaling $g_{\mu \nu} \to 
\Lambda^2 g_{\mu \nu}$ the Einstein term can
be put into canonical form.
At leading order $\Lambda$ can be replaced by 
$\vol$ and one obtains the Weyl rescaled low energy
effective Lagrangian
(we set $\kappa_{11}= 1$ henceforth)\footnote{As we said
before $ {\cal L}^{(3)}_0$ is also corrected at this order
but we did not compute those correction.}
\be\label{M3d3}
{\cal L}^{(3)}\  =\   {\cal L}^{(3)}_0
- \frac{1}{2}\,\epsilon^{\mu \nu \rho}\, \tilde W_{AB} 
 A^A_\mu F^B_{\nu \rho}
\ -\, V \ ,
\ee
where ${\cal L}^{(3)}_0$ is given in (\ref{M3d}) and one has 
\beqn \label{defs}
V  & = &  \frac{1}{4\vol^{3}}\left(
\int_{Y_4} d^8 \xi \sqrt{g^{(8)}} |F_4|^{2} 
- \frac{1}{6} T_2\, \chi\right)\ ,\nonumber\\
\tilde W_{AB} &=& \frac{1}{2} \partial_A\partial_B \tilde W\ ,\qquad
\tilde W = \frac14\int_{Y_4} F_4 \wedge J\wedge J\ . 
\eeqn
For $\chi=0$ both $V$ and the Chern-Simons terms were absent.\footnote{
The fact that the $E_8$-term contributes
to the potential in $D=3$ was first noticed in \cite{AFMN}.}

In order to display the relationship of the potential 
with the two  superpotentials of \cite{GVW} we need to
further rewrite $V$. By definition we have 
\be\label{Fstar}
\int_{Y_4} d^8 \xi \sqrt{g^{(8)}} |F_4|^{2} =  
\int_{Y_4} F_4 \wedge \star F_4 \ ,
\ee
where to leading order $\star F_4$ is the Hodge dual of $F_4$
with respect to the metric $g_{mn}$.
$F_4$ can be expanded as the sum 
$F_4 = F_{4,0} + F_{3,1} + F_{2,2} 
+ F_{1,3} + F_{0,4}$.
In order to proceed let us recall that on $Y_4$
a primitive $(p,q)$-form $\omega^{(0)}_{p,q}$
can be defined by
\begin{equation}\label{primitive}
J^{5-p-q} \wedge \omega^{(0)}_{p,q} = 0\ ,
\end{equation} 
where $J^{n}$ is the $n$th wedge product of
the K\"ahler form.
The Hodge dual of a primitive four-form
is given by  \cite{GVW}
\begin{equation}\label{starprim}
\star \omega^{(0)}_{p,4-p} = (-1)^p \omega^{(0)}_{p,4-p}\ .
\end{equation}
On $Y_4$ all components of $F_4$  except $F_{(2,2)}$
are primitive in that they satisfy (\ref{primitive})
and as a consequence their Hodge duals are  simply given 
by
\be \label{pdual}
\star F_{4,0} = F_{4,0}\ , \qquad
\star F_{3,1} = - F_{3,1}\ , \qquad
\star F_{1,3} = - F_{1,3}\ , \qquad
\star F_{0,4} = F_{0,4}\ .
\ee
For $F_{(2,2)}$ one uses 
the Lefschetz decomposition which asserts   \cite{GH}
\be 
F_{2,2} \equiv 
F^{(0)}_{2,2} + J \wedge F^{(0)}_{1,1} + J^2 \wedge F^{(0)}_{0,0}\ ,
\ee
and computes explicitly \cite{MH} 
\be\label{dualp}
\star F_{2,2} = 
F^{(0)}_{2,2} - J \wedge F^{(0)}_{1,1} + J^2 \wedge F^{(0)}_{0,0} = F_{2,2} - 2J \wedge F^{(0)}_{1,1}\ .
\ee
Combining eqs.\ (\ref{pdual})--(\ref{dualp})
one arrives at
\begin{equation}\label{starg}
\star F_4 = F_4 - 2 F_{3,1}   - 2 F_{1,3} 
- 2 J \wedge F^{(0)}_{1,1}  \ ,
\end{equation}
and hence
\begin{equation}\label{gg}
\int_{Y_4} F_4 \wedge \star F_4 = \int_{Y_4} F_4 \wedge F_4  
- 4 \int_{Y_4} F_{3,1} \wedge F_{1,3}
- 2 \int_{Y_4} J \wedge F^{(0)}_{1,1} \wedge J \wedge F^{(0)}_{1,1}\ ,
\end{equation}
where we have used $J\wedge F^{(0)}_{2,2} = 0$
and $J^3 \wedge F^{(0)}_{1,1} = 0$ 
in accord with (\ref{primitive}).

The second term in (\ref{gg}) can be further rewritten
by using \cite{AS}
\be\label{domega}
D_\alpha\Omega \equiv  \partial_\alpha \Omega
+ (\partial_\alpha K_{3,1})\, \Omega = \Phi_\alpha
\ee
where $\Phi_\alpha$ is the basis of $H^{3,1}$
and $K_{3,1}$ is the K\"ahler potential for the $(3,1)$ moduli 
defined in eq.\ (\ref{Galphabeta}).
With the help of (\ref{domega}) and (\ref{Galphabeta})
one derives
\be\label{almost}
\int_{Y_4} F_{3,1} \wedge F_{1,3}
= - e^{K_{3,1}} 
G^{-1 \alpha\bar \beta} D_\alpha W D_{\bar \beta} \bar W \ ,
\ee
where 
\begin{equation} \label{wtildew}
W = \int_{Y_4} \Omega \wedge F_4\ 
\end{equation}
is precisely the chiral superpotential of \cite{GVW}.

Finally, with the help of (\ref{GAB})
also the last term in (\ref{gg}) can be expressed
in terms of the superpotential  (\ref{defs}).
Expanding $F^{(0)}_{1,1}$ into the basis $e_A$ and using
\be \label{g-1}
G^{-1 AB} = -\frac{1}{6} \vol \vol^{-1 AB} + \frac{2}{3} M^A M^B\ ,
\ee
where $\vol_{AB} = \frac{1}{12} \partial_A \partial_B \vol$ and indices are raised with
$\delta^{AB}$, one finds 
\be\label{final}
\int_{Y_4} J \wedge F^{(0)}_{1,1} \wedge J \wedge F^{(0)}_{1,1}
= - \vol^{-1} \left( G^{-1 AB} \partial_A \tw \partial_B \tw 
- 2 \tw^2\right)\ .
\ee
Inserting (\ref{almost}), (\ref{final}) into  (\ref{gg})
and using (\ref{Fstar}) one arrives at 
\beqn \label{fsquare}
\int_{Y_4} d^8 \xi \sqrt{g^{(8)}} |F_4|^{2} 
& = & \int_{Y_4} F_4 
\wedge F_4 + 4 e^{K_{3,1}} 
G^{-1 \alpha\bar \beta} D_\alpha W D_{\bar \beta} \bar W  \non
&&\quad + 2 \ \vol^{-1} \left( G^{-1 AB} \partial_A \tw \partial_B \tw 
- 2 \tw^2\right)\ .
\eeqn
Inserting (\ref{fsquare}) into (\ref{defs}) and 
taking into account the tadpole cancellation condition 
(\ref{tadpol}) 
results in  
\be \label{r4d3action5}
V= e^{K^{(3)}} 
G^{-1 \alpha\bar \beta} D_\alpha W D_{\bar \beta} \bar W 
+
\vol^{-4} \left(\frac{1}{2} G^{-1 AB} \partial_A \tw \partial_B \tw 
- \tw^2  \right) \ , 
\ee
where
\be \label{k3}
K^{(3)} = K_{3,1} -3\ln \vol\ .
\ee

Finally, $V$ and ${\cal L}^{(3)}$ can be written in a more
canonical form by transforming to new coordinates 
\begin{equation} \label{mhat}
\hat M^A = \vol^{-1} M^A\ , \qquad
\hat J = \hat M^A e_A\ .
\end{equation}
{}From eqs.\ (\ref{GAB}), (\ref{eqvolumekf}) and (\ref{k3}) 
we learn
\begin{equation} \label{vhat}
\hat \vol \equiv \int_{Y_4}
\hat J^4
= \vol^{-3}
\ , \qquad \hat G_{AB} = \vol^2 G_{AB}
= -\frac{1}{2} \partial_A \partial_B \ln \hat \vol
\ ,
\qquad
K^{(3)} = K_{3,1} + \ln \hat \vol \ , 
\end{equation}
where the derivatives $\partial_A$ are now with respect to $\hat M^A$.
If we furthermore introduce 
\begin{equation} \label{hatw}
\hat W = \frac{1}{4} \int_{Y_4}  F_4\wedge\hat J \wedge \hat J 
= \vol^{-2} \tw
\end{equation}
and insert   (\ref{vhat})  and 
(\ref{hatw}) into (\ref{M3d3}) using (\ref{M3d}) we arrive at
\beqn\label{M3d4}
{\cal L}^{(3)}  &=&  \frac{1}{2} R^{(3)} 
- G_{\alpha \bar{\beta}} \pu Z^\alpha 
\po \bar{Z}^{\bar{\beta}} 
  -\, \frac{1}{2} \hat G_{AB} 
\pu \hat M^A \po \hat M^B 
 - \frac{1}{4}  \hat G_{AB} 
F^A_{\mu \nu} F^{B \mu \nu} \nonumber \\
&& - \frac{1}{2}\,
\epsilon^{\mu \nu \rho}\, \hat W_{AB} 
 A^A_\mu F^B_{\nu \rho}\ -\, V \ ,
\eeqn
where the explicit dependence on $\hat \vol$ has 
disappeared from the Lagrangian.
The potential in the new variables is given by
\begin{equation} \label{Potential2}
V = e^{K^{(3)}} G^{-1\alpha \bar \beta} D_\alpha W 
D_{\bar \beta} \bar W + 
\frac{1}{2} \hat G^{-1 AB} \partial_A \hat W
\partial_{B} \hat W - \hat W^2\ ,
\end{equation}
where the derivatives $\partial_A \hat W$  are again taken
with respect to the new variables $\hat M^A$.

We see that the potential is entirely expressed
in terms of two superpotentials $W$ and $\hat W$.
The $W$ given in (\ref{wtildew})
is precisely the chiral superpotential of \cite{GVW} while
$\hat W$ given in (\ref{hatw})
is the real version of the superpotential
of \cite{GVW}. This is related to the fact that the presence of the 
Chern-Simons terms no longer allows a duality transformation
from vector to chiral multiplets. 
As a consequence $\hat W$ can not be complexified
as there are only real scalars $\hat M^A$
in the vector multiplets. However, upon
further $S^1$ reduction $\hat W$ should become complex
and coincide with the $D=2$ superpotential of \cite{G,GVW}.\footnote{%
To show this directly would require a 
generalization of the arguments given in \cite{W2}. The reduction of the
Chern-Simons terms leads to terms $\sim \theta_A F^A_{01}$ and the
$\theta$-angles contribute to the potential in $D=2$.}

Finally, let us compare the potential (\ref{Potential2})
with the potentials of $D=3, N=2$ supergravity.
Unfortunately, the relevant potential for chiral and vector
multiplets with Chern-Simons terms
coupled to $D=3,N=2$ supergravity is not available
in the literature. The full derivation of this potential
is beyond the scope of this paper and will be presented
elsewhere.
Here we just derive part of it by an $S^1$ compactification
of a corresponding $D=4$ supergravity. 

The four-dimensional theory we need to consider
has to contain  both chiral and linear multiplets.
A $D=4$ linear multiplet consists of an antisymmetric tensor
and a real scalar $L$ as bosonic components.
The $D=4$ Lagrangian is determined
by two functions,
the holomorphic superpotential $W(\phi^\alpha)$
and the real function 
$K^{(4)} = 
K_\phi (\phi^\alpha, \bar \phi^{\bar \alpha}) + K_L (L^{\tilde A})$,
where $\phi^\alpha$ denote the scalars of the chiral multiplets.
$K_\phi$ is the K\"ahler potential of the chiral fields
and the second derivative of $K_L$ determines the
$\sigma$-model metric of $L^{\tilde A}$  \cite{BGG,DQQ}.\footnote{%
We have chosen a theory where $K^{(4)}$ is the sum
of two terms
since this matches the situation we have in $D=3$.}
In this theory the scalar potential is given by\footnote{We
thank R.\ Grimm for discussions on this point.} 
\begin{equation} \label{potential}
V^{(4)} = e^{K^{(4)}} \left( G^{-1\alpha \bar \beta} D_\alpha W 
D_{\bar \beta} \bar W 
+ (L^{\tilde A} \partial_{\tilde A} K^{(4)} - 3) |W|^2 \right)\ ,
\end{equation}
where $D_\alpha W = 
\partial_\alpha W +(\partial_\alpha K^{(4)}) W$.
%
This form of the potential can be derived for example 
{} from the $D=4$ duality between an antisymmetric tensor 
and a scalar. At the level of superfields  this results
in the  duality between 
a linear multiplet $L$ and a chiral multiplet $S$
with $S+\bar S \sim L^{-1}$. The K\"ahler potential 
of $S$ is given by  $K=-\ln(S+\bar S)$ and the superpotential
continues to be a function of only 
the $\phi^\alpha$. In this dual description
with only chiral multiplets $V^{(4)}$
takes the standard form
$V^{(4)} = e^{K^{(4)}} \left( G^{-1I \bar J} D_I W 
D_{\bar J} \bar W - 3 |W|^2 \right)$,
where the index $I$ now runs over all chiral
multiplets $(\phi^\alpha,S^{\tilde A})$.

Reducing the theory on a circle leaves the chiral multiplets
unaltered. The linear
multiplets, however,  become vector multiplets and  an additional 
vector multiplet containing the radius $r$  and the Kaluza-Klein 
vector of the circle as its bosonic components appears 
in the spectrum. Let us define $L^0 = r^{-2}$.
A straightforward $S^1$-reduction shows that after an
appropriate Weyl rescaling the $D=3$ potential is given by
\begin{equation} \label{potential2}
V^{(3)} = e^{K^{(3)}} \left( G^{-1\alpha \bar \beta} D_\alpha W D_{\bar \beta} \bar W 
+ (L^A \partial_A K^{(3)} - 4) |W|^2 \right)\ ,
\end{equation}
where
$K^{(3)} = K^{(4)} + \ln L^0$ 
and the index $A$ now includes 0.
This form of the potential is indeed consistent with the first term of
(\ref{Potential2}) if one identifies $L^A=\hat M^A$,
uses (\ref{vhat}) and the identity 
$\hat M^A\partial_A \ln\hat\vol = 4$.

The second term in (\ref{Potential2}) is precisely the D-term
of a $D=3, N=2$ Chern-Simons theory coupled to supergravity.
As was noted in ref.\ \cite{NG} supersymmetrization
of the Chern-Simons terms requires a D-term potential 
which coincides with the second term in (\ref{Potential2}).
The last term  of (\ref{Potential2})
should arise when not only a pure  Chern-Simons theory 
but a more general gauge theory including
the standard kinetic term and the Chern-Simons term
is coupled to $N=2$ supergravity. 

Curiously,  the potential for $\hat W$
as displayed in (\ref{Potential2})
is the $D=3$ version of a general formula  for the 
scalar potential in arbitrary dimension $D$.
The requirement of the stability of AdS backgrounds 
leads to \cite{PT}
\begin{equation} \label{potential3}
V^{(D)} = \frac12(D-2)(D-1)\left( \frac{D-2}{D-1} G^{-1 AB} \partial_A
\hat W \partial_{B} \hat W - \hat W^2 \right)\ ,
\end{equation}
which for $D=3$ indeed gives $V^{(3)} =
\frac{1}{2} G^{-1 AB} \partial_A \hat W
\partial_{B} \hat W - \hat W^2$.

Finally, let us discuss
the conditions for unbroken supersymmetry. 
{}From our derivation of the $D=3$ potential
it is clear that for the chiral multiplets
unbroken supersymmetry requires $D_\alpha W = 0$
which in turn imposes constraints on the allowed $4$-form 
flux $F_4$.
Using (\ref{wtildew})  $D_\alpha W = 0 = D_{\bar \beta} \bar W $ 
implies $F_{3,1}= F_{1,3}=0$.
Furthermore, from our derivation of the potential 
one would expect that for the vector multiplets
the conditions of unbroken supersymmetry are\footnote{A 
rigorous derivation requires the computation
of the fermionic supersymmetry transformations
in an $N=2$ supergravity background.}
\begin{equation} \label{susy}
(\partial_A K^{(3)})\, W\ =\ 0\ =\ 
\partial_A \hat W \ , 
\end{equation}
while the vacuum energy is determined by  $\hat W^2$ only.
The first condition is the precise analog of the four-dimensional
situation when the superpotential does not depend
on a chiral scalar, in which case 
$D_S W = (\partial_S K) W$ holds and supersymmetry also forces
$W=0$. As we have seen the $D=3$ vector multiplets
are closely related to the $D=4$ linear multiplets
which precisely have the feature $D_S W = (\partial_S K) W$.
$(\partial_A K^{(3)}) W=0$ in general can only be
fulfilled if $W=0$ which implies 
 $F_{4,0}= F_{0,4}=0$.
Finally, using (\ref{hatw})
$\partial_A \hat W = 0$ implies $\hat J\wedge F_4 =0$,
i.e.\ $F_4$
has to be primitive. 
This last condition implies not 
only $\partial_A \hat W = 0$ but also $\hat W = 0$. 
Thus the cosmological constant always vanishes
in a supersymmetric minimum and no supersymmetric $AdS_3$
solution exists.%
\footnote{Recently this has also been noticed
in the revised version of \cite{GVW}. 
We thank the authors for communicating their result prior 
to publication.}
To summarize, supersymmetry constrains 
the $4$-form flux to be a primitive $(2,2)$-form
$F_4 = F_{2,2}^{(0)}$. This result 
has first been derived in \cite{BB} and has led to the proposal 
of the two superpotentials
(\ref{wtildew}) and (a complex version of) (\ref{hatw}) in ref.\ \cite{GVW}. 
By performing a Kaluza-Klein reduction on a Calabi-Yau fourfold 
with $4$-form flux turned on we have verified that 
to order ${\cal O} (\kappa_{11}^{-2/3})$ 
the potential can 
be expressed in terms of 
the superpotentials (\ref{wtildew}), (\ref{hatw})
and reproduced the conditions for a supersymmetric ground state.

\vskip 0.5cm

\noindent
{{\bf Acknowledgments}}

We thank R.\ Grimm,
S.\ Gukov, H.\ G\"unther, C.\ Herrmann, M.\ Klein, 
A.\ Klemm, T.\ Mohaupt
and M.\ Zagermann for valuable discussions. 
We also thank
R.\ Grimm for his hospitality at the CPT in Marseille
where part of this work was done.
J.L.\ thanks D.\ Gross and S.\ Weinberg 
for financial support and hospitality
at the ITP, Santa Barbara and the University of Texas in Austin
at the final stages of this work.

This research was supported in part by the
DFG (the German Science Foundation), 
GIF (the German--Israeli 
Foundation for Scientific Research), the
DAAD (the German Academic Exchange Service) and
the National Science Foundation under Grant No. PHY99-07949.


\end{document}